\documentclass[twocolumn,english,aps,prb,floatfix,preprintnumbers,showpacs,amsfonts,amssymb]{revtex4}
\usepackage[T1]{fontenc}
\usepackage[latin1]{inputenc}
\usepackage{amsmath}
\usepackage{amssymb}

\makeatletter

\providecommand{\LyX}{L\kern-.1667em\lower.25em\hbox{Y}\kern-.125emX\@}

\usepackage{graphicx}
\usepackage{graphicx}
\usepackage{graphicx}
\usepackage{amscd}
\usepackage{bm}

\usepackage{babel}
\makeatother
\begin{document}

\title{Coupled quantum wires}
\newcommand{\be}{\begin{equation}}
\newcommand{\ee}{\end{equation}}
\newcommand{\bd}{\begin{displaymath}}
\newcommand{\ed}{\end{displaymath}}

\author{D.\ Makogon, N.\ de Jeu, and C.\ Morais Smith}

\affiliation{Institute for Theoretical Physics, University of
Utrecht, Leuvenlaan 4, 3584 CE Utrecht, The Netherlands.}

\date{\today{}}

\begin{abstract}
We study a set of crossed 1D systems, which are coupled with each
other via tunnelling at the crossings. We begin with the simplest
case with no electron-electron interactions and find that besides
the expected level splitting, bound states can emerge. Next, we
include an external potential and electron-electron interactions,
which are treated within the Hartree approximation. Then, we write
down a formal general solution to the problem, giving additional
details for the case of a symmetric external potential.
Concentrating on the case of a single crossing, we were able to
explain recent experinents on crossed metallic and semiconducting
nanotubes [J.\ W.\ Janssen, S.\ G.\ Lemay, L.\ P.\ Kouwenhoven,
and C.\ Dekker, Phys.\ Rev.\ B \textbf{65}, 115423 (2002)], which
showed the presence of localized states in the region of crossing.
\end{abstract}


\maketitle
\section{Introduction}
Physics in 1D systems manifests a number of peculiar phenomena,
such as spin-charge separation, conductance
quantization,\cite{BLandauer} and anomalous low-temperature
behavior in the presence of backscattering impurity.\cite{Kane1}
It is reasonable to expect that the more complex structures
composed of crossed 1D systems, such as crossings and arrays,
should exhibit some particular features as well. Although the
transport properties of crossed 1D systems and their arrays have
been thoroughly studied both theoretically\cite{Komnik} and
experimentally\cite{Gao,Fuhrer,Postma}, the electronic structure
of these systems is much less understood and the interpretation of
existing experimental results is challenging. Recent scanning
tunnelling microscopy (STM) experiments on a metallic carbon
nanotube crossed with a semiconducting one\cite{Janssen} have
shown the existence of localized states at
the crossing which are not 
due to disorder. However, these localized states do not appear
systematically in all experiments, i.e. the effect is highly
dependent on the nature of the carbon nanotubes (metallic or
semiconducting), of the barrier formed at the crossing, etc.
Aiming at clarifying this problem, we present in this paper a
detailed study of tunnelling effects between crossed 1D systems in
the presence of potential barriers for massive quasiparticle
excitations. Because effects of electron-electron interactions can
be reasonably incorporated in a random phase approximation
(RPA),\cite{DzLarkin, DasSarma} we study a simpler model,
accounting for electron-electron interactions only within Hartree
approximation. The outline of this paper is the following: in
section II we introduce the model that we are going to use to
describe the array of crossed nanowires. In section III we
consider a particular case of free electrons and write down
explicit solutions for the case of one and four crossings. Section
IV contains formal general solution with additional details given
for the case of a symmetric external potential. We demonstrate the
effect of tunnelling on the electronic structure of single
crossings in Section V and qualitatively discuss different
possibilities depending on the external potential. Section VI
contains quantitative analysis and comparison with available
experimental data of the electronic structure of single crossing
for different values of parameters. Our conclusions and open
questions are presented in Section VII.
\section{The Model}
We consider a system composed of two layers of crossed quantum
wires with interlayer coupling. The upper layer has a set of
parallel horizontal wires described by fermionic fields
$\psi_{j}(x)$, whereas the lower layer contains only vertical
parallel wires described by the fields $\varphi_{i}(y)$. The wires
cross at the points $(x_i,y_j)$, with $i,j \in Z$ and the distance
between
layers is $d$, with $min(|x_i-x_{i+1}|,|y_j-y_{j+1}|)\gg d$, see Fig.1.\\
\begin{figure}[htb]
\begin{center}
\includegraphics[width=6cm,angle=0]{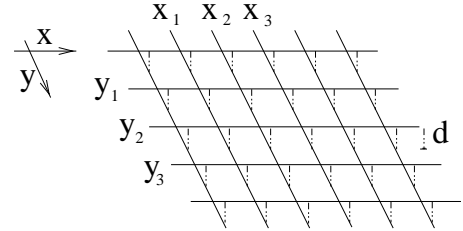}
\end{center}
\caption{2D array of crossed wires. }
\end{figure}\\
The partition function of the system reads
\begin{equation}
Z=\int d[\psi_j] d[\psi^*_j] d[\varphi_{i}]
d[\varphi_{i}^*]e^{-S/\hbar},
\end{equation}
with the total action given by
 \begin{equation}
S=S_0+S_{\rm sct}+S_{\rm int}.\label{Taction}
\end{equation}
The first term accounts for the kinetic energy and external
 potential $V^{\rm ext}_{j}(x)$, which can be different in each wire and may arise, e.g.,
due to a lattice deformation, when one wire is built on top of
another,
\begin{eqnarray}\nonumber
 S_{0}&=&\sum_j\int_0^{\hbar \beta}d\tau\int dx\psi_{j}^*
(x,\tau)G^{-1}_{jx}\psi_{j}(x,\tau)\\
&+& \sum_i\int_0^{\hbar\beta}d\tau\int dy\varphi_{i}^*
(y,\tau)G^{-1}_{iy}\varphi_{i}(y,\tau),\label{action0}
\end{eqnarray}
where
\begin{eqnarray}\nonumber
G^{-1}_{jx}&=&\hbar\frac{\partial}{\partial\tau}-\frac{\hbar^2}{2
m}\frac{d^2}{d x^2}+V^{\rm ext}_{j}(x)-\mu_x,\\
G^{-1}_{iy}&=&\hbar\frac{\partial}{\partial\tau}-\frac{\hbar^2}{2
m}\frac{d^2}{d y^2}+V^{\rm ext}_{i}(y)-\mu_y. \label{Greenf}
\end{eqnarray}
Here, $\mu_{x,y}$ denotes the chemical potential in the upper
($\mu_{x}$) or lower ($\mu_{y}$) layer.

The second term of Eq.\ (\ref{Taction}) describes scattering at
the crossings $(x_i,y_j)$,
 \begin{equation}
S_{\rm sct}=\sum_{ij}\int_0^{\hbar\beta}d\tau H_{ij},
\end{equation}
where
\begin{equation}\nonumber
H_{ij}=\left[ \psi_{j}^*
(x_i,\tau) \quad \varphi^*_{i}(y_j,\tau) \right]\left( \begin{array}{cc} U_{ij} & T_{ij}\\
 T^*_{ij} &  \tilde{U}_{ij}
\end{array}
 \right)
 \left[ \begin{array}{c} \psi_{j}(x_i,\tau)\\ \varphi_{i}(y_j,\tau)
\end{array}
 \right].
\end{equation}
Notice that the matrix element $U_{ij}$ describing intra-layer
contact scattering can, in principle, be different from
$\tilde{U}_{ij}$, but both must be real. On the other hand, the
contact tunnelling (inter-layer) coefficient between the two
crossed wires $T_{ij}$ can be a complex number, since the only
constraint is that the matrix above must be Hermitian.

 \widetext
 The third term in Eq.\ (\ref{Taction}) accounts for electron-electron
interactions,
\begin{eqnarray}\nonumber
 S_{\rm int}&=&\frac{1}{2}\sum_j\int_0^{\hbar\beta}d\tau\int_0^{\hbar\beta}d\tau'\int dx\int dx'\psi_{j}^*
(x,\tau)\psi_{j}^* (x',\tau')V^{\rm
e-e}(x-x')\psi_{j}(x,\tau)\psi_{j}(x',\tau')\\
&+&
\frac{1}{2}\sum_i\int_0^{\hbar\beta}d\tau\int_0^{\hbar\beta}d\tau'\int
dy\int dy'\varphi_{i}^* (y,\tau)\varphi_{i}^* (y',\tau')V^{\rm
e-e}(y-y')\varphi_{i}(y,\tau)\varphi_{i}(y',\tau').\label{actionint}
\end{eqnarray}
\endwidetext
\section{Free electrons case}
We start by considering  a very simplified case, namely, free
electrons (no electron-electron interaction, $V^{\rm e-e}(x)=0$
and no external potential, $V^{\rm ext}_{j}(x)=0$). Moreover, we
assume $\tilde{U}_{ji}=U_{ji}=0$ and put $\mu_x=\mu_y=\mu$. The
interlayer tunnelling is assumed to be equal at each crossing
point $T_{ij}=T$ and to have a real and positive value. In such a
case, the partition function consists of only Gaussian integrals.
We can then integrate out the quantum fluctuations, which reduces
the problem to just solving the equations of motion. Considering a
real time evolution and performing a Fourier transformation in the
time variable, we are left with the following equations of motion
for the fields:
\begin{eqnarray}\nonumber
 \left(-\frac{\hbar^2}{2 m}\frac{d^2}{d x^2}-E\right)\psi_{j}(x)+T\sum_{l} \delta(x-x_l)\varphi_{l}(y_j)&=&
 0,\\
\left(-\frac{\hbar^2}{2 m}\frac{d^2}{d
y^2}-E\right)\varphi_{i}(x)+T\sum_{l}
\delta(y-y_l)\psi_{l}(x_i)&=&
 0,\label{EOM1}
\end{eqnarray}
where $m$ denotes the electron mass and $E$ is the energy of an
electron state. Firstly, we evaluate the solutions for the case of
free electrons without tunnelling and then we investigate how the
addition of tunnelling changes the results. The solution for the
free electron case consists of symmetric and antisymmetric
normalized modes,
\begin{equation}
\psi_{s}(x)=\frac{1}{\sqrt{L}}\cos(k_s x), \qquad
\psi_{a}(x)=\frac{1}{\sqrt{L}}\sin(k_a x), \label{freesol}
\end{equation}
respectively. The corresponding momenta $k_s$ and $k_a$ depend on
the boundary conditions: with open boundary conditions $k_s=\pi
(2n+1)/2L$, $k_a=\pi n/L$ and with periodic boundary conditions
$k_s=k_a=\pi n/L$ for a wire of length $2L$ and $n$ integer. To
find the solution for the case with tunnelling $T\neq0$, we have
to solve Eqs.\ (\ref{EOM1}). These equations are linear,
therefore, the solution consists of a homogeneous and an
inhomogeneous parts,
\begin{equation}
\psi_{j}(x)=\psi_{j}^{\rm hom}(x)+\psi_{j}^{\rm inh}(x),
\end{equation} which are
\begin{equation}
\psi_{j}^{\rm hom}(x)=A_j e^{i k x}+B_j e^{-i k x}, \end{equation}
\begin{equation}
\psi_{j}^{\rm inh}(x)=\frac{T m}{\hbar^2 k
}\sum_{l}\varphi_{l}(y_j)\sin(k|x-x_l|). \label{Inh1}
\end{equation} Imposing open boundary conditions, $\psi_{j}(\pm
L)=0$, we find
\begin{eqnarray}\nonumber
 A_j e^{i k L}+B_j e^{-i k L}+\psi_{j}^{\rm inh}(L)&=&
 0,\\
A_j e^{-i k L}+B_j e^{i k L}+\psi_{j}^{\rm inh}(-L)&=&
 0.\label{EOM2}
\end{eqnarray}
Writing the above equations in a matrix notation and inverting
yields
\begin{equation}\nonumber
\left( \begin{array}{c}  A_j\\  B_j
\end{array}
 \right)=\frac{-1}{2 i \sin(2 k L)}\left( \begin{array}{cc} e^{i k L} & -e^{-i k L}\\
 -e^{-i k L} &  e^{i k L}
\end{array}
 \right)
 \left( \begin{array}{c} \psi_{j}^{\rm inh}(L)\\ \psi_{j}^{\rm inh}(-L)
\end{array}
 \right).
\end{equation}
Substituting explicitly the expression for $\psi_{j}^{\rm inh}(\pm
L)$ given by Eq.\ (\ref{Inh1}) and using the mathematical identity
\begin{eqnarray}\nonumber
\left( e^{i k x} \quad e^{-i k x} \right)\left( \begin{array}{cc} e^{i k L} & -e^{-i k L}\\
 -e^{-i k L} &  e^{i k L}
\end{array}
 \right)
 \left( \begin{array}{c} \sin(kL-kx_l)\\ \sin(kL+kx_l)
\end{array}
 \right)\\ \nonumber
=\cos\left(2 k L\right)\cos(kx-kx_l)-\cos(kx+kx_l),
\end{eqnarray}
leads, after simplifications, to the solution
\begin{eqnarray}\nonumber
 \psi_{j}(x)&=&-T\sum_{l} G(x,x_l)\varphi_{l}(y_j),\\
 \varphi_{i}(y)&=&-T\sum_{l} G(y,y_l)\psi_{l}(x_i),\label{FSol1}
\end{eqnarray}
where, for open boundary conditions,
\begin{eqnarray}\nonumber
 G_o(x_i,x_j,E)&\equiv&\frac{m}{\hbar^2 k\sin(2 k L)}
[\cos(kx_i+kx_j)\\  &-&\cos(2kL-k|x_i-x_j|)],\label{SimpleRep}
\end{eqnarray}
and the energy $E$ is related to $k$ as $E=\hbar^2 k^2/2m$.
Similar calculations can be performed for the case of periodic
boundary conditions, yielding Eq.\ (\ref{FSol1}) with
\begin{equation}
 G_p(x_i,x_j,E)\equiv\frac{m}{\hbar^2 k\sin(k
L)}\cos(kL-k|x_i-x_j|).\label{PeriodRep}
\end{equation}
\subsection{Two crossed wires}
In particular, for the simplest case of a single horizontal and a
single vertical wires, with just one crossing at $(x_0,y_0)$, the
solution is:
\begin{eqnarray}\nonumber
 \psi(x)&=&-T G(x,x_0,E)\varphi(y_0)\\
 \varphi(y)&=&-T G(y,y_0,E)\psi(x_0).\label{FSol2}
\end{eqnarray}
\begin{figure}[htb]
\begin{center}
\includegraphics[width=6cm,angle=0]{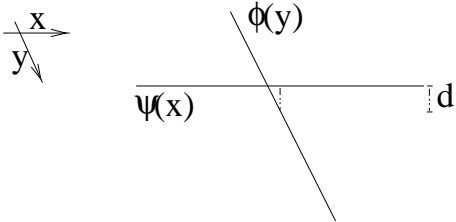}
\end{center}
\caption{Two crossed wires.}
\end{figure}\\
By substituting $(x,y)=(x_0,y_0)$, we find that at the crossing
point
\begin{eqnarray}\nonumber
 \psi(x_0)&=&-T G(x_0,x_0,E)\varphi(y_0)\\
 \varphi(y_0)&=&-T G(y_0,y_0,E)\psi(x_0).\label{FSol3}
\end{eqnarray}
The consistency condition requires that
\begin{equation}
\left| \begin{array}{cc} 1 & T G(x_0,x_0,E)\\
 T G(y_0,y_0,E) &  1
\end{array}
 \right|=0,
\end{equation}
or
\begin{equation}
T^2 G(x_0,x_0,E) G(y_0,y_0,E)=1.\label{FreeSpectr}
\end{equation}
The solution is even simpler if $(x_0,y_0)=(0,0)$. Then, for open
boundary conditions, the symmetric modes are
\begin{eqnarray}\nonumber
 \psi(x)&=&\frac{\varphi(0)T m}{\hbar^2 k \cos(k
 L)}\sin(k L-k|x|), \\ \nonumber
 \varphi(y)&=&\frac{\psi(0)T m}{\hbar^2 k \cos(k
 L)}\sin(k L-k|y|),
\end{eqnarray}
and the antisymmetric modes are left unchanged in comparison with
Eqs.\ (\ref{freesol}). Also,
\begin{equation}
 G(0,0,E)=\frac{m \tan(k L)}{\hbar^2k}, \label{nonintfs}
\end{equation} and the secular equation
(\ref{FreeSpectr}) becomes
\begin{equation}
\left[\frac{T m\tan(k L)}{\hbar^2 k}\right]^2=1,
\end{equation}
which splits into two transcendental equations
\begin{eqnarray}\nonumber
 k^+&=&-\frac{T m}{\hbar^2}\tan(k^+ L), \\ \nonumber
 k^-&=&\frac{T m}{\hbar^2}\tan(k^- L).
\end{eqnarray}
The first one describes the shifted values of scattering states
energies, whereas the second equation has an additional bound
state solution with $E<0$, if $T>T_0=\hbar^2/mL$. The appearance
of the bound state is exclusively due to the presence of
tunnelling. For an electron in a wire of length $2L=10^3$ nm the
corresponding value is $T_0=7.62\times 10^{-5}$ eV$\cdot$nm and
for quasiparticles the value of $T_0$ is typically larger,
inversely proportional to their effective mass. Defining then
$\kappa\equiv-i k^-$ and taking the thermodynamic limit
$L\rightarrow \infty$, we find $|\kappa|=T m/\hbar^2$ with the
corresponding bound state energy
\begin{equation} E=-\frac{T^2 m}{2\hbar^2},\label{BEnergy}
\end{equation}
and the wave function given by
\begin{equation}
\psi(x)=\frac{\sqrt{|\kappa|}}{2}e^{-|\kappa x|}.\label{Expsol}
\end{equation}
The factor $1/2$ instead of $1/\sqrt{2}$ comes from the fact that
now an electron can tunnel into the other wire, where its
wavefunction $\varphi(0)=-\psi(0)$.
 Eqs.\ (\ref{BEnergy}) and (\ref{Expsol}) hold for both
open and periodic boundary conditions. Since the threshold value
$T_0$ is quite small, the bound state should exist for a typical
crossing with relatively good contact. However, the energy of the
state is extremely small, $E\sim 10^{-8}$ eV if $T\sim T_0$.
Qualitatively similar results were found by numerical
computation\cite{Schult, Carini} of the ground-state energy of an
electron trapped at the intersection of a cross formed by two
quantum wires of finite width.
\subsection{Four crossed wires}
 For the case of two wires in the upper and two in the lower layers, there are four
crossings. In this case, the self consistent equations read
\begin{equation}
\left[ \begin{array}{c} \psi_{1}(x_1)\\
 \psi_{1}(x_2)\\ \psi_{2}(x_1)\\
 \psi_{2}(x_2)
\end{array}
 \right]=M(x_1,x_2,E)
 \left[ \begin{array}{c} \varphi_{1}(y_1)\\
 \varphi_{1}(y_2)\\ \varphi_{2}(y_1)\\
 \varphi_{2}(y_2)
\end{array}
 \right]
\end{equation}
and
\begin{equation}
 \left[ \begin{array}{c} \varphi_{1}(y_1)\\
 \varphi_{1}(y_2)\\ \varphi_{2}(y_1)\\
 \varphi_{2}(y_2)
\end{array}
 \right]=M(y_1,y_2,E)\left[ \begin{array}{c} \psi_{1}(x_1)\\
 \psi_{1}(x_2)\\ \psi_{2}(x_1)\\
 \psi_{2}(x_2)
\end{array}
 \right],
\end{equation}
\widetext where
\begin{equation}
M(x_1,x_2,E)=-T\left( \begin{array}{cccc}  G(x_1,x_1,E) & 0 &  G(x_1,x_2,E) & 0\\
  G(x_1,x_2,E) &  0 &  G(x_2,x_2,E) & 0\\
 0 &  G(x_1,x_1,E) & 0 &  G(x_1,x_2,E)\\
  0 &  G(x_1,x_2,E) & 0 &  G(x_2,x_2,E)
\end{array}
 \right).
\end{equation}
\endwidetext
The secular equation then has the form
\begin{equation}
\det[M(x_1,x_2,E)M(y_1,y_2,E)-I]=0,\label{SpectralEq}
\end{equation}
which yields a rather complicated transcendental equation ($I$ is
the identity matrix). The spectral equation for bound states $E<0$
can be significantly simplified in the thermodynamic limit
$L\rightarrow \infty$. Then, with $k=i \kappa$, for both open and
periodic boundary conditions, the matrix elements become
\begin{equation}
 G(x_i,x_j,E)=\frac{m}{\hbar^2 |\kappa|}
e^{-|\kappa(x_i-x_j)|}\label{fundbound}
\end{equation}
and the secular equation in Eq.\ (\ref{SpectralEq}) has 4
solutions with negative energy described by
\begin{eqnarray}\nonumber
E&=&-\frac{T^2 m}{2\hbar^2}(1-a_1-a_2+a_1 a_2),\\ \nonumber
E&=&-\frac{T^2 m}{2\hbar^2}(1+a_1-a_2-a_1 a_2),\\ \nonumber
E&=&-\frac{T^2 m}{2\hbar^2}(1-a_1+a_2-a_1 a_2),\\ \nonumber
E&=&-\frac{T^2 m}{2\hbar^2}(1+a_1+a_2+a_1 a_2). \nonumber
\end{eqnarray}
Here, $a_1\equiv e^{-|\kappa(x_2-x_1)|}$, $a_2\equiv
e^{-|\kappa(y_2-y_1)|}$, and $E=-\hbar^2 \kappa^2/2m$ (notice the
implicit dependence of $a_1$ and $a_2$ on $E$). The value of $a_i$
depends exponentially on the distance between the crossing points.
In the limit $|x_2-x_1|,|y_2-y_1|\rightarrow \infty$ the value of
$a_1,a_2\rightarrow 0$, which correspond to four independent
crossings with the bound state energy $E=-T^2 m/{2\hbar^2}$, the
same value as we found in the previous case (see Eq.\
(\ref{BEnergy})).
\subsection{A regular lattice of crossed wires}
Consider now a regular square lattice, with lattice constant $a$.
Then, one has $x_l=al$ and $y_j=aj$. From symmetry arguments, the
wave functions should be $\psi_{j}(x)=\psi_{0}(x)e^{iK_y aj}$ and
$\varphi_{l}(y)=\varphi_{0}(y)e^{iK_x al}$. After substituting
them into Eq.\ (\ref{FSol1}) and using Eq.\ (\ref{fundbound}) we
find
\begin{eqnarray}\nonumber \psi_j(x)=&-&T\varphi_0(y_j)\frac{me^{i K_x l_x
a}}{\hbar^2 \kappa}\left[\frac{\sinh(\kappa x -
\kappa a l_x)e^{i K_x a }}{\cosh(\kappa a )-\cos(K_x a )}\right.\\
\nonumber  &-&\left.\frac{\sinh(\kappa x - \kappa
(l_x+1)a)}{\cosh(\kappa a )-\cos(K_x a )}\right],\\ \nonumber
\varphi_l(y)=&-&T\psi_0(x_l)\frac{me^{i K_y l_y a}}{\hbar^2
\kappa}\left[\frac{\sinh(\kappa x -
\kappa a l_y)e^{i K_y a }}{\cosh(\kappa a )-\cos(K_y a )}\right.\\
\nonumber  &-&\left.\frac{\sinh(\kappa y - \kappa
(l_y+1)a)}{\cosh(\kappa a )-\cos(K_y a )}\right],
\end{eqnarray}
where  $l_x,l_y \in Z$, such that $al_x\leq x<a(l_x+1)$ and
$al_y\leq y<a(l_y+1)$. Therefore,
$\psi_{j}(x_l)=\psi_{0}(0)e^{i(K_x al+K_y aj)}$ and
$\varphi_{l}(y_j)=\varphi_{0}(0)e^{i(K_x al+K_y aj)}$, with
$\psi_{0}(0)$ and $\varphi_{0}(0)$ related by
\begin{eqnarray}\nonumber
 \psi_0(0)&=&-T\frac{m}{\hbar^2 \kappa}\frac{\sinh(\kappa
a )}{\cosh(\kappa a )-\cos(K_x a )}\;\varphi_0(0),\\
 \varphi_0(0)&=&-T\frac{m}{\hbar^2 \kappa}\frac{\sinh(\kappa
a )}{\cosh(\kappa a )-\cos(K_y a )}\;\psi_0(0).\label{BandSol}
\end{eqnarray}
Thus, the spectral equation reads
\begin{equation}\nonumber
 1=\frac{(mT)^2}{(\hbar^2 \kappa)^2}\frac{\sinh^2(\kappa
a )}{[\cosh(\kappa a )-\cos(K_x a )][\cosh(\kappa a )-\cos(K_y a
)]}.
\end{equation}
By performing an analytic continuation $k=i \kappa$ in Eq.\
(\ref{BandSol}), we find an equations similar to the one obtained
previously by Kazymyrenko and Dou\c{c}ot\cite{Kazymyrenko} when
studying scattering states in a lattice. The spectral equation
describes a band formed by bound states with energies $-T/a<E<0$.
The momenta $K_x$ and $K_y$ run in the interval
$-\pi<K_xa,K_ya<\pi$ if $T\geq T_f=2\hbar^2/ma$ or inside the
region $|\sin(K_xa/2)\sin(K_ya/2)|\leq T/T_f$ if $T<T_f$. Similar
results were calculated,\cite{Dickinson}
estimated,\cite{CastroNeto} and measured\cite{Zhou} in the context
of hybridization between vertical and horizontal stripe modes in
high-Tc superconductors.
\section{A more general case}\
Now we consider a more general model, which takes into account the
presence of an inhomogeneous potential $V^{\rm ext}_{j}(x)$
arising from possible lattice deformations, and includes
electron-electron interactions $V^{\rm e-e}(x)$, which will be
treated at a mean field level, within the Hartree approximation
$V^{\rm e-e}_{{\rm H} j}(x)$. Each crossing $(x_i,y_j)$ is
considered as a scattering point with tunnelling $T_{ij}$ and
scattering potential $U_{ij}$. The corresponding equations of
motion then read
\begin{eqnarray}\nonumber
 D_{jx}\psi_{j}(x)+\sum_{l}[U_{lj}\psi_{j}(x_{l})+T_{lj}\varphi_{l}(y_j)]\delta(x-x_l)&=&
 0,\\\nonumber
D_{iy}\varphi_{i}(x)+\sum_{l}[\tilde{U}_{il}\;\varphi_{i}(y_l)+T^*_{il}\psi_{l}(x_{i})
]\delta(y-y_l)&=&
 0,
\end{eqnarray}
where
\begin{eqnarray}\nonumber
 D_{jx}&=&-\frac{\hbar^2}{2 m}\frac{d^2}{d x^2}+V_{j}(x)-E,\\ \nonumber
D_{iy}&=&-\frac{\hbar^2}{2 m}\frac{d^2}{d
y^2}+V_{i}(y)-E,
\end{eqnarray}
with $V_{j}(x)=V^{\rm ext}_{j}(x)+V^{\rm e-e}_{{\rm H} j}(x)$.
This model is solved most easily through the Green's function
satisfying
\begin{equation} \nonumber
D_{jx_1} G_j(x_1,x_2,E)=\delta (x_1-x_2)
\end{equation}
with
\begin{equation} \nonumber
 G_j(x_1,x_2,E)= G_j^*(x_2,x_1,E),
\end{equation}
and the corresponding open boundary conditions,
\begin{equation}\nonumber
  G_j(x_1,L,E)=0,\quad  G_j(x_1,-L,E)=0,
\end{equation}
or the periodic ones
\begin{eqnarray}\nonumber
  G_j(x_1,L,E)= G_j(x_1,-L,E),\\ \nonumber
{ G_j}'(x_1,L,E)={ G_j}'(x_1,-L,E),
\end{eqnarray}
where the prime denotes the derivative with respect to $x_1$. Note
that we consider real time Green's function for a particular wire
(not the whole system). The solution to the model is
\begin{eqnarray}\nonumber
 \psi_{j}(x)&=&-\sum_{l}[U_{lj}\psi_{j}(x_{l})+T_{lj}\varphi_{l}(y_j)] G_j(x,x_l,E),\\
\varphi_{i}(y)&=&-\sum_{l}[\tilde{U}_{il}\;\varphi_{i}(y_l)+T^*_{il}\psi_{l}(x_{i})
] G_i(y,y_l,E),\label{GenSol}
\end{eqnarray}
which we require to be normalized
\begin{equation}
\sum_{l}\left(\int |\psi_{l}(x)|^2 dx+\int
|\varphi_{l}(y)|^2dy\right)=1.
\end{equation}
The self consistency condition for the value of the functions at
crossing points $(x_i,y_j)$ yields the equations
\begin{eqnarray}\nonumber
\sum_{l}[(U_{lj}
G_j(x_i,x_l,E)+\delta_{il})\psi_{j}(x_{l})\\\nonumber +T_{lj}
G_j(x_i,x_l,E)\varphi_{l}(y_j)]&=&0,\\\nonumber
\sum_{l}[(\tilde{U}_{il}\; G_s(y_j,y_l,E)+\delta_{jl})\varphi_{i}(y_l)\\
+T^*_{il} G_i(y_j,y_l,E)\psi_{j}(x_{i}) ]&=&0.\label{Cons}
\end{eqnarray}
To find nontrivial solutions for the fields $\psi_{j}(x)$ and
$\varphi_{i}(y)$, the system of homogeneous equations in Eq.\
(\ref{Cons}) has to be linearly dependent and hence the solution
is represented by the null space of the system. This means that
after writing the equations in a matrix form, the determinant of
the matrix should be zero, thus leading to a spectral equation for
$E$. 
Moreover, bound state solutions in the thermodynamic limit
$L\rightarrow \infty$ satisfy both open and periodic boundary
conditions, since $\psi(\pm L)\rightarrow 0$ and $\psi'(\pm
L)\rightarrow 0$.

\begin{figure}[htb]
\begin{center}
\includegraphics[width=6cm,angle=0]{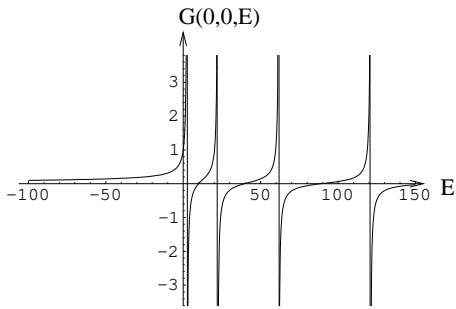}
\end{center}
\caption{\label{fig3}$ G(0,0,E)$ in units of $m/\hbar^2$ versus
$E$ in units of $\hbar^2/2mL^2$. }
\end{figure}

To understand better the dependence of the Green's function $
G_j(x_i,x_l,E)$ on $E$, we represent the function through the
solutions of the homogenous equations,
\begin{equation}
D_{jx}\psi_j(x)=0.
\end{equation}
We omit the index $j$ in what follows for simplicity. The most
general and common representation, which holds for any static
potential, reads as follows:
\begin{equation}
 G(x_1,x_2,E)=\sum_{n}\frac{\psi_{\varepsilon_{n}}^*(x_1)\psi_{\varepsilon_{n}}(x_2)}{\varepsilon_{n}-E}.
\label{fsrep0}
\end{equation}
Here, the function $\psi_{\varepsilon}(x)$ is the solution of the
homogenous equation
\begin{equation}
\left(-\frac{\hbar^2}{2 m}\frac{d^2}{d
x^2}+V(x)-\varepsilon\right)\psi_\varepsilon(x)=0,
\end{equation}
and the spectrum $\{\varepsilon_{n}\}$ is obtained by imposing the
corresponding boundary conditions. 
Notice that in the present representation of $ G(x_1,x_2,E)$ the
functions $\psi_{\varepsilon_{n}}(x)$ have to be orthonormal. By
writing $ G(x_1,x_2,E)$ in the form given in Eq.\ (\ref{fsrep0}),
the following identity arises
\begin{equation}
\int dx'
 G(x_1,x',E) G(x',x_2,E)=\frac{\partial G(x_1,x_2,E)}{\partial
E}.
\end{equation}
The case $x_1=x_2=0$ for free electrons is illustrated in Fig.\
\ref{fig3}, where Eq.\ (\ref{nonintfs}) is plotted. If some
external potential is present, $ G(x_0,x_0,E)$ has the same form
but the positions of the poles are shifted and the corresponding
values are different.
If no regularization is used, the calculations for $E>0$ must be
performed in the finite size limit, otherwise with
$L\rightarrow\infty$ the energy distance between different modes
vanishes and the poles situated on the real positive half axis
merge to form a branch cut singularity. This behavior can be
readily seen on the example of Eq.\ (\ref{nonintfs}), where can
perform an analytic continuation, considering $k\rightarrow k+
ik'$. Then, in the limit $L\rightarrow\infty$, $\tan(kL + ik'L)=i
{\rm sgn}(k')$, and the function $ G(x_0,x_0)$ changes sign as one
goes from the upper to the lower complex half plane for $k\neq0$.

Now we represent the Green's function through the solutions of the
homogenous equation
\begin{equation}
\left(-\frac{\hbar^2}{2 m}\frac{d^2}{d
x^2}+V(x)-E\right)\psi(x)=0.  \label{Srdng}
\end{equation}
This is a second order differential equation, therefore, it should
have two linearly independent solutions, which we call $\psi_1(x)$
and $\psi_2(x)$. Then the Green's function is
\begin{equation}
 G(x_1,x_2,E)=\left\{\begin{array}{c} A_{-}\psi_1(x_1)+B_{-}\psi_2(x_1), x_1\leq x_2\\
A_{+}\psi_1(x_1)+B_{+}\psi_2(x_1), x_1 >x_2
\end{array} \right.
 \label{fsrepf},
\end{equation}
where the expressions for the coefficients
$A_{-},B_{-},A_{+},B_{+}$ (functions of $x_2$), are derived in the
Appendix A. In particular, for a symmetric potential $V(x)$, we
can choose a symmetric $\psi_s(x)$ and an antisymmetric
$\psi_a(x)$ solutions as linearly independent, i.e.,
$\psi_1(x)=\psi_s(x)$ and $\psi_2(x)=\psi_a(x)$. Thus we find
\begin{equation}
 G(x,0,E)=\frac{m\psi_{a}(L)}{\hbar^2{\psi_{a}}'(0)}\left[\frac{\psi_{s}(x)}{\psi_{s}(L)}-\frac{\psi_{a}(|x|)}{\psi_{a}(L)}\right]
\end{equation}
and
\begin{equation}
 G(0,0,E)=\frac{m\psi_{s}(0)}{\hbar^2{\psi_{a}}'(0)}\frac{\psi_{a}(L)}{\psi_{s}(L)}.
\label{fs00n}
\end{equation}
To obtain the results in the thermodynamic limit $L\rightarrow
\infty$, it is useful to rewrite $ G(x_1,x_2)$ using quantities
which do not depend on $L$ explicitly. For example,
\begin{equation}
 G(x,0,E)= G(0,0,E)\frac{\psi_{s}(x)}{\psi_{s}(0)}-\frac{m}{\hbar^2}\frac{\psi_{a}(|x|)}{{\psi_{a}}'(0)}.
\end{equation}
After substitution of Eqs.\ (\ref{freesol}) into Eq.\
(\ref{fsrepf}) and simplification, for the case of noninteracting
electrons we find
\begin{eqnarray}\nonumber
 G(x_1,x_2,E)&=&\frac{m}{\hbar^2k\sin(2 k L)}
[\cos(kx_1+kx_2)\\
\nonumber &-&\cos(2kL-k|x_1-x_2|)],\nonumber
\end{eqnarray}
which is the same expression as in the previous section (see Eq.\
(\ref{SimpleRep})). This is a posteriori justification of the use
of the same letter $ G(x_1,x_2,E)$ in the first section. The case
of a harmonic potential is considered in Appendix B.
\section{a single crossing}
Now we apply our results including tunnelling and external
potential 
to the simpler case of only two crossed wires, aiming to compare
our findings with experiments. Using the general solution given by
Eq.\ (\ref{GenSol}), and considering $T=T^*$, we can write
\begin{eqnarray}\nonumber
 \psi(x)&=&-[U\psi(x_{0})+T\varphi(y_0)] G_1(x,x_0,E),\\
\nonumber \varphi(y)&=&-[\tilde{U}\;\varphi(y_0)+T\psi(x_{0}) ]
G_2(y,y_0,E).\label{Solsingle}
\end{eqnarray}
By substituting $(x,y)=(x_0,y_0)$, we find that at the crossing
point
\begin{eqnarray}\nonumber
 [1+U G_1(x_0,x_0,E)]\psi(x_{0})+T G_1(x_0,x_0,E)\varphi(y_0)&=&0,\\
\nonumber [1+\tilde{U} G_2(y_0,y_0,E)]\varphi(y_0)+T
G_2(y_0,y_0,E)\psi(x_{0})&=&0.\label{Solsinglecross}
\end{eqnarray}
The consistency condition requires that
\begin{equation}
\left| \begin{array}{cc} 1+U G_1(x_0,x_0,E) & T G_1(x_0,x_0,E)\\
 T G_2(y_0,y_0,E) &  1+\tilde{U} G_2(y_0,y_0,E)
\end{array}
 \right|=0,
\end{equation}
or
\begin{eqnarray}\nonumber
0&=&[1+U G_1(x_0,x_0,E)][1+\tilde{U} G_2(y_0,y_0,E)] \\
\nonumber &-&T^2 G_1(x_0,x_0,E) G_2(y_0,y_0,E).
\end{eqnarray}
The meaning of this equation becomes clearer in the symmetric
case, when $U=\tilde{U}$ and $ G_1(x_0,x_0,E)= G_2(y_0,y_0,E)= G$.
In this case, it reduces to a quadratic equation, which bears two
solutions,
\begin{equation}\nonumber
 G_+=\frac{-1}{U+T},\qquad  G_-=\frac{-1}{U-T}.
\end{equation}
Notice that they differ by the sign in front of the tunnelling
amplitude $T$, which is shifting the potential $U$. Such symmetry
effectively reduces the problem to 1D with effective potential
$U_{\rm eff}\delta(x_0)$. Hence, we have
\begin{eqnarray}\nonumber
 \psi(x_{0})&=&\varphi(y_0),\quad \; \; U_{\rm eff}^+=U+T,\\
\psi(x_{0})&=&-\varphi(y_0),\quad U_{\rm
eff}^-=U-T.\label{SymSolsingle}
\end{eqnarray}
The shift of the energy levels in a wire due to the presence of
the $\delta$ potential can be visualized with the help of the
Green's function expansion, where one has
\begin{equation}
 G(x_0,x_0,E)=\sum_{n}\frac{|\psi_{\varepsilon_{n}}(x_0)|^2}{\varepsilon_{n}-E}=\frac{-1}{U_{\rm
eff}}.
\end{equation}
In the case with $U_{\rm eff}=0$, the energies are exactly those
of the poles and, therefore, remain unshifted. However, since $
G(x_0,x_0,E)=-1/U_{\rm eff}$, the curve actually describes how the
energies of the modes change as we keep increasing $-1/U_{\rm
eff}$ from $-\infty$ if $U_{\rm eff}>0$ or decreasing $-1/U_{\rm
eff}$ from $+\infty$ if $U_{\rm eff}<0$. In the latter case, we
can run into the region with $E<0$, which would correspond to the
appearance of a bound state. Nevertheless, to obtain an exact
solution, it is more convenient to work with the expression for $
G(x_0,x_0,E)$ in terms of the wave functions,
\begin{equation}
 G(0,0,E)=\frac{m\psi_{s}(0)}{\hbar^2{\psi_{a}}'(0)}\frac{\psi_{a}(L)}{\psi_{s}(L)}=\frac{-1}{U_{\rm
eff}},
\end{equation}
where we assumed $x_0=0$ for simplicity.

\section{comparison with experiments}

\begin{figure}[htb]
\begin{center}
\includegraphics[width=6cm,angle=0]{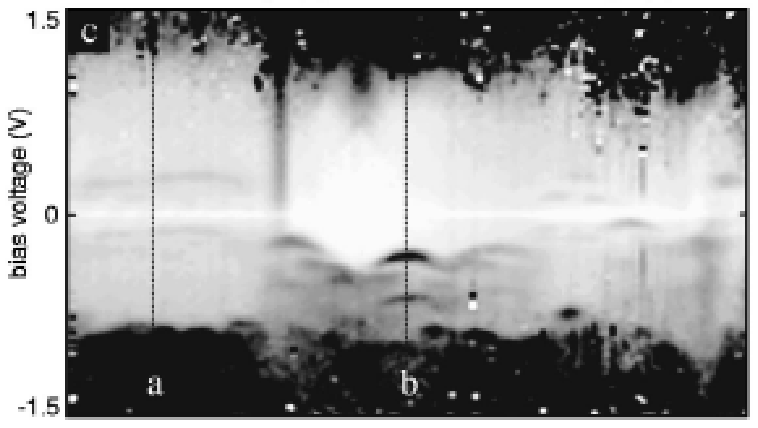}
\end{center}
\caption{\label{figexp} Voltage versus length diagram, which shows the
experimentally observed density of states. Notice the existence of two
localized states in black. (Extracted from Ref.\ \onlinecite{Janssen}).}
\end{figure}

Now, we will compare our theoretical findings with experimental
results. We concentrate mostly on the analysis of a system
consisting of two crossed single wall carbon nanotubes (SWNTs): a
metallic on top of a semiconducting (MS) one.\cite{Janssen} In its
unperturbed state, the band structure of a SWNT can be understood
by considering the electronic structure of graphene. Due to its
cylindrical shape, the transverse momentum of one particle
excitations in a SWNT has to be quantized, whereas the
longitudinal momentum may vary continuously. Combining this
condition with the assumption that the electronic structure is not
very different from that of graphene, one finds two different
situations, depending on the topology of the SWNT: there are no
gapless modes and the nanotube is semiconducting, or two gapless
modes are present and the nanotube is called metallic. Analyzing
the spectroscopic measurements performed along the metallic
nanotube (see Fig.\ \ref{figexp}) and comparing with the unperturbed
electronic structure,
one notices two main changes. First, a small quasi gap opens
around the Fermi energy level $\varepsilon_F$ between
$\varepsilon_F-0.2$ eV and $\varepsilon_F+0.3$ eV in the spectrum
of the massless modes (corresponding to zero transverse momentum).
Second, two peaks are visible at $\varepsilon_0=\varepsilon_F-0.3$
eV and $\varepsilon_1=\varepsilon_F-0.6$ eV in the region around
the crossing, corresponding to localized states between the Fermi
energy and the van Hove singularity at $\varepsilon_{\rm vH
}=\varepsilon_F-0.8$ eV. Such states are not visible above the
Fermi energy, thus suggesting that the electron-hole symmetry is
broken by the presence of some external potential. The latter may
appear due to lattice distortions and the formation of a Schottky
barrier at the contact between the nanotubes.
\cite{Odintsov,OdintsovYo} In the following, we show that if the
potential is strong enough, localized states can form in the
spectrum of the massive mode corresponding to the van Hove
singularity with energy $\varepsilon=\varepsilon_{\rm vH }-E$.
Therefore, the observed localized states should have $E_0=-0.5$ eV
and $E_1=-0.2$ eV.

To incorporate in a more complete way the effects of the Schottky
barrier and lattice deformation, we assume $V^{\rm ext}(x)$ to
have a Lorentzian shape,
\begin{equation}
V^{\rm ext}(x)=-\frac{\tilde{V}}{1+x^2/b^2}.\label{Potential}
\end{equation}

\begin{figure}[htb]
\begin{center}
\includegraphics[width=6cm,angle=0]{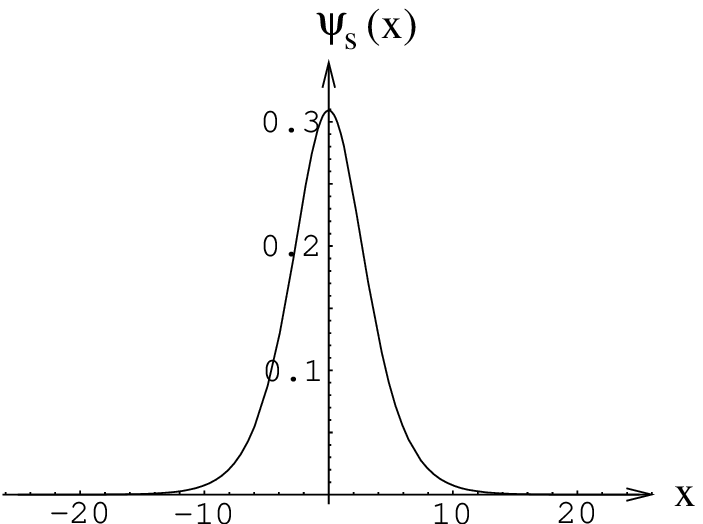}
\end{center}
\caption{\label{fig4}$\psi_s(x)$(nm$^{-1/2}$) versus $x$ (nm) }
\end{figure}

\begin{figure}[htb]
\begin{center}
\includegraphics[width=6cm,angle=0]{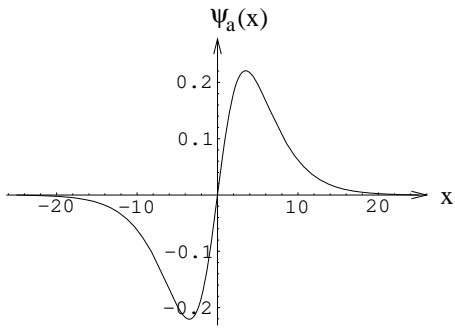}
\end{center}
\caption{\label{fig5}$\psi_a(x)$(nm$^{-1/2}$) versus $x$ (nm). }
\end{figure}

Firstly, we study the influence of this potential alone on the
electronic structure, i.e. we assume that there is no tunnelling
$T=0$, and no electron-electron interactions. Exact numerical
solution of the Schrodinger equation shows that an approximation
of the potential in Eq.\ (\ref{Potential}) by the harmonic one
does not change the solution qualitatively. Therefore, we consider
$V^{\rm ext}(x)\approx-\tilde{V}(1-x^2/b^2)$, which describes a
harmonic oscillator with frequency
$\omega=\sqrt{2\tilde{V}/mb^{2}}$ and corresponding spectra
$E_{n}=-\tilde{V}+(n+1/2)\sqrt{2\hbar^{2}\tilde{V}/mb^{2}}$ for
$E_{n}<0$. Moreover, it is reasonable to assume that the strength
of the barrier $\tilde{V}$ is of the same order as the energy of
the bound states and that the potential is localized on the same
length scale as the localized states. Hence, we take
$\tilde{V}=0.7$ eV and $b=4$ nm. It follows then from our
calculations that the difference between neighboring energy levels
is quite small and there are many bound states present in the case
when $m$ is the actual electron mass. However, assuming $m$ to be
an effective electron mass, with $m=0.025 \; m_e$, which is of the
same order as the experimentally estimated values $m=0.037 \;
m_e$\cite{JarilloHerrero} and $m=0.06 \; m_e$,\cite{Radosavljevic}
we find exactly two pronounced bound states: the first one has
$E=-0.5$ eV and is described by the symmetric wavefunction
$\psi_s(x)$ as shown in Fig.\ \ref{fig4},
whereas the other has $E=-0.2$ eV and is described by the
antisymmetric wavefunction $\psi_a(x)$, see Fig.\ \ref{fig5}.
Considering Fig.\ \ref{fig4}, we observe that the localization
size of the state is around $10$ nm, which agrees well with the
experimental data. On the other hand, the state shown in Fig.\
\ref{fig5} has a zero value exactly at the crossing and is rather
spread, a behavior which is not observed experimentally. Besides
these two, a number of other states are also present in the
vicinity of the van Hove singularity with $E>-0.1$ eV.

\begin{figure}[htb]
\begin{center}
\includegraphics[width=6cm,angle=0]{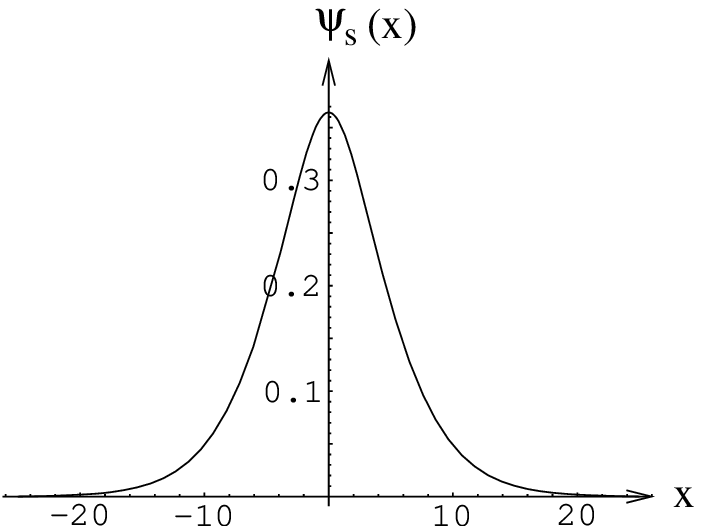}
\end{center}
\caption{\label{fig4int}$\psi_s(x)$(nm$^{-1/2}$) versus $x$ (nm) }
\end{figure}

Secondly, we take into account electron-electron interactions to
consider other possibilities to obtain two pronounced bound
states. Unfortunately, our approach only allows us to incorporate
electron-electron interactions at the mean-field level by using
the Hartree selfconsistent approximation
\begin{equation}
V^{\rm e-e}_H(x)=\int dx' V^{\rm e-e}(x-x')n(x'),
\end{equation}
where $n(x)$ is the electron density, given by
\begin{equation}
n(x)=\sum_k |\psi_k(x)|^2 n_F(\varepsilon_k-\mu).
\end{equation}
Here the summation $k$ goes over energy levels and
$n_F(\varepsilon)$ is the Fermi distribution. Although it is known
that in 1D systems quantum fluctuations play an extremely
important role, we nevertheless start with the mean-field
approximation as a first step to incorporate them in RPA.
Moreover, we believe that their presence does not qualitatively
change the obtained results. To render the numerical calculation
simpler, we consider a delta-like interaction potential, which
leads to
\begin{equation}
V^{\rm e-e}_H(x)=V_0 n(x),
\end{equation}
By estimating the effective interaction strength $V_0\sim 2\pi
\hbar v_F$ from the Luttinger liquid theory, we obtain that
$V_0\sim 3.4$ eV$\cdot$nm for $v_F=8.2\times 10^7$
cm/s.\cite{Lemay} Suppose that the lowest energy state with
$E=-0.5$ eV is occupied by an electron with a certain spin. Then,
there is a possibility to add to the same state an electron with
an opposite spin. However, due to the repulsive Coulomb
interaction the energy of the two-electron state becomes $E=-0.2$
eV for $V_0=3.15$ eV$\cdot$nm. The corresponding self consistent
solution is presented in Fig.\ \ref{fig4int}.
The state has the same shape as in Fig.\ \ref{fig4}, but is a bit
more spread. By comparing the density of states (DOS) distribution
with scanning tunnelling spectroscopy (STS) data for the
crossing,\cite{Janssen} we observe that the inclusion of
electron-electron interactions (Fig.\ \ref{fig4int}) provides a
much better agreement between theory and experiment for the
$E=-0.2$ eV bound state than in the previous case (Fig.\
\ref{fig5}).

\begin{figure}[htb]
\begin{center}
\includegraphics[width=6cm,angle=0]{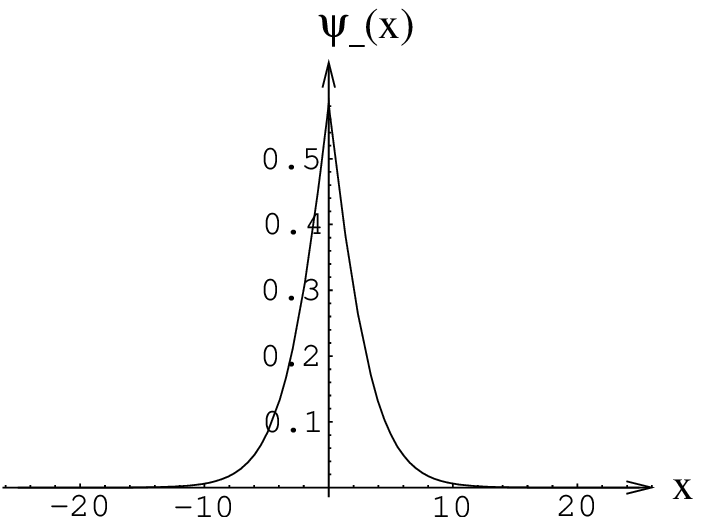}
\end{center}
\caption{\label{fig6}$\psi_-(x)$(nm$^{-1/2}$) versus $x$ (nm). }
\end{figure}

\begin{figure}[htb]
\begin{center}
\includegraphics[width=6cm,angle=0]{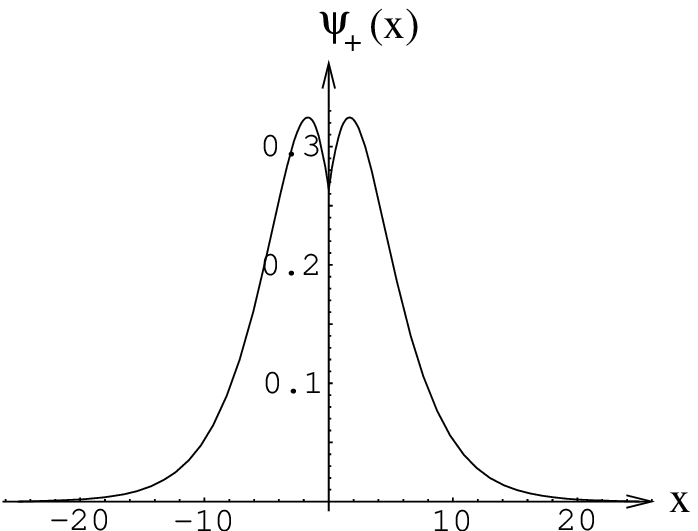}
\end{center}
\caption{\label{fig7}$\psi_+(x)$(nm$^{-1/2}$) versus $x$ (nm). }
\end{figure}

\begin{figure}[htb]
\begin{center}
\includegraphics[width=6cm,angle=0]{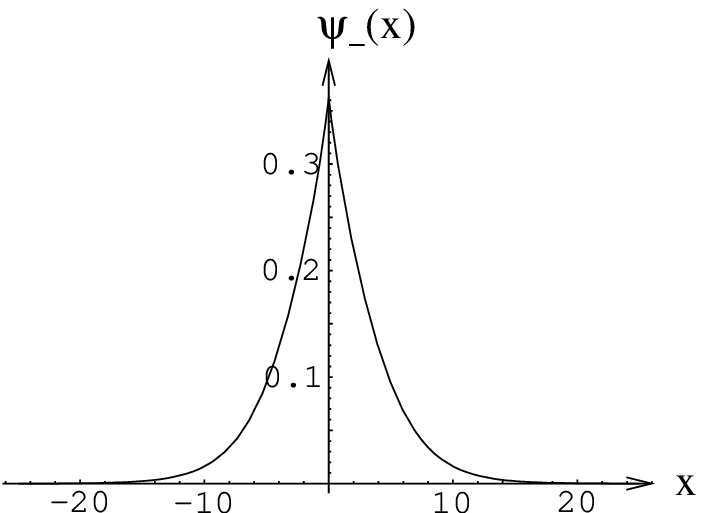}
\end{center}
\caption{\label{fig8}$\psi_-(x)$(nm$^{-1/2}$) versus $x$ (nm). }
\end{figure}

Thirdly, we take into account tunnelling between the wires.
Qualitatively, this leads to the splitting of energy levels and
redistribution of charge density in the wires, thus effectively
reducing the strength of electron-electron interactions. Since we
have no information about the electronic structure of the
semiconducting nanotube, to make a quantitative estimation we
assume that the effective mass is equal in both wires and that the
potential is also the same. In such a case, from symmetry arguments the
electron density should be evenly distributed in both wires even
for a very weak tunnelling. Therefore, the electron-electron
interactions should be twice stronger than in the case without
tunnelling, namely, $V_0=6.3$ eV$\cdot$nm to achieve the same
energy value. Moreover, if the tunnelling coefficient is large
enough, the splitting of the energy levels becomes significant and
detectable. We can estimate the coefficient $T$, if we assume that
it has the same order for SM, metallic-metallic (MM), and
semiconducting-semiconducting (SS) nanotube junctions. The SS and
MM junctions have Ohmic voltage-current dependance, characterized
by the conductance $G$. Moreover, we can estimate the transmission
coefficient of the junction as $G/G_0 \sim (T/2\pi \hbar
v_F)^{2}$, for $G/G_0\ll 1$.  For MM junctions experimental
measurements\cite{Dickinson} typically yield $G/G_0\sim 10^{-2}$,
thus corresponding to $T\sim 0.34$ eV$\cdot$nm. For example, for
$T=0.28$ eV$\cdot$nm and $\tilde{V}=0.44$ eV in Eq.\
(\ref{Potential}), without electron-electron interactions
 we find that the system has {\it two bound states}.
The lowest energy bound state with $E=-0.5$ eV is shown in Fig.\
\ref{fig6}.
Compared with Fig.\ \ref{fig4}, the state has a peak exactly at
the crossing, corresponding to a local increase of the DOS. The
other bound state with $E=-0.2$ eV is shown in Fig.\ \ref{fig7}.
Contrary to the previous case, the state has a deep at the
crossing, corresponding to a local decrease of the DOS. However,
these local change in DOS is too small to be observable in the
present experimental data. If we now include electron-electron
interactions with $V_0=3.15$ eV$\cdot$nm and add a second electron
with different spin to the system, we find that the new state has
$E=-0.267$ eV and acquires the shape shown in Fig.\ \ref{fig8}.

The last result suggests that there are yet other possible
interpretations of the experimental results. Firstly, if the
potential in the metallic SWNT is significantly decreased due to
screening effects but a Schottky barrier in the semiconducting
SWNT can reach considerable values, sufficient for the formation
of the bound states, then the latter are also going to be present
in the metallic SWNT due to tunnelling between SWNTs. Secondly,
there is still a possibility to find a bound state existing purely
due to tunnelling, i.e., without external potential, as was shown
in Eq.\ (\ref{Expsol}), and a second bound state may arise with
different energy due to Coulumb repulsion between electrons with
different spins. However, this is most probably not the case we
have in the experiments, because due to electron-hole symmetry
such states would exist also above the Fermi energy, a result
which is not observed experimentally.
\section{Conclusions}
We presented several possibilities to explain the observed
localized states at the crossing of metallic and semiconducting
nanotubes.\cite{Janssen} All of them require the existence of an
external potential in the metallic and/or semiconducting SWNT to
break the electron-hole symmetry, since the localized states were
seen only below the Fermi energy. Most probably, such a potential
comes from a Schottky barrier and the effect of lattice
distortions is minimal, since such localized states were, up to
now, observed only for MS crossings and not for MM or SS ones.
Moreover, the effective mass of quasiparticle excitations should
be of order $m=0.025 \; m_e$, where $m_e$ is the actual electron
mass, to generate only a few bound states localized on a region of
approximately $10$ nm with energy of order of $0.5$ eV. The best
agreement with the experimental data is obtained by assuming that
the second bound state has a different energy due to the Coulumb
repulsion between electrons with different spins. The role of
tunnelling in the observed electronic structure is not clear and
allows for many interpretations. To avoid such ambiguity, the
electronic structure of the semiconducting nanotube should be
measured as well. Moreover, to be sure that the available STS
measurements indeed represent the electronic structure of the
nanotube and are free of artifacts introduced by the STM tip
\cite{LeRoy} several measurements with different tip height should
be performed.

\section{ACKNOWLEDGMENTS}
We are very grateful to S.\ G.\ Lemay for useful discussions.

\appendix
\section{}
Here we consider the Green's function as a function of one
variable $x_1$ and fix $x_2$ for a moment. Since $ G(x_1,x_2,E)$
is the Green's function, we require it to satisfy proper boundary
conditions $ G(\pm L,x_2,E)=0$, be continuous $ G(x_2-0,x_2,E)=
G(x_2+0,x_2,E)$, and also $ G'(x_2-0,x_2,E)-
G'(x_2+0,x_2,E)=2m/\hbar^2$. Substituting Eq.\ (\ref{fsrepf}) into
the above requirements one finds
\begin{equation}
P \left( \begin{array}{c} A_+\\
 A_-\\ B_+\\
 B_-
\end{array}
 \right)= \frac{2 m}{\hbar^2}\left( \begin{array}{c} 0\\
 0\\ 0\\
 1
\end{array}
 \right),\label{Acoef}
\end{equation}
where
\begin{equation}
P\equiv\left( \begin{array}{cccc} \psi_{1}(L) & 0 & \psi_{2}(L) & 0\\
 0 &  \psi_{1}(-L) & 0 & \psi_{2}(-L)\\
 \psi_{1}(x_2) & -\psi_{1}(x_2) & \psi_{2}(x_2) & -\psi_{2}(x_2)\\
 -\psi_{1}'(x_2) & \psi_{1}'(x_2) & -\psi_{2}'(x_2) & \psi_{2}'(x_2)
\end{array}
 \right).
\end{equation}
Multiplying the Eq.\ (\ref{Acoef}) by the matrix $P^{-1}$ we find
\begin{equation} \nonumber
\left( \begin{array}{c} A_+\\
 A_-\\ B_+\\
 B_-
\end{array}
 \right)= C \left( \begin{array}{c} \psi_{2}(L)[\psi_{2}(-L)\psi_{1}(x_2)-\psi_{1}(-L)\psi_{2}(x_2)]\\
 \psi_{2}(-L)[\psi_{2}(L)\psi_{1}(x_2)-\psi_{1}(L)\psi_{2}(x_2)]\\
  -\psi_{1}(L)[\psi_{2}(-L)\psi_{1}(x_2)-\psi_{1}(-L)\psi_{2}(x_2)]\\
 -\psi_{1}(-L)[\psi_{2}(L)\psi_{1}(x_2)-\psi_{1}(L)\psi_{2}(x_2)]
\end{array}
 \right),
\end{equation}
where
\begin{equation} \nonumber
C\equiv\frac{2m}{\hbar^2
W_r}[\psi_{1}(L)\psi_{2}(-L)-\psi_{1}(-L)\psi_{2}(L)]^{-1}.
\end{equation}
The Wronskian 
\begin{equation}\nonumber
W_r\equiv \psi_{1}(x_2)\psi_{2}'(x_2)-\psi_{2}(x_2)\psi_{1}'(x_2),
\end{equation}
is nonzero for linearly independent functions and its value does
not depends on the point $x_2$.

\section{}
Suppose that Eq.\ (\ref{Srdng}) has a solution $\psi(x)$ which is
neither symmetric nor antisymmetric. Thus, for symmetric
potentials $\psi(-x)$ is also a solution and both of them are
linearly independent. Furthermore, we can then compose a symmetric
$\psi_s(x)=(\psi(x)+\psi(-x))/2$ and an antisymmetric
$\psi_a(x)=(\psi(x)-\psi(-x))/2$ solutions. In particular, for a
harmonic potential $V(x)=m \omega^2 x^2/2$, one can find such a
solution
\begin{equation}
\psi(x)=e^{-\frac{m \omega x^2}{2\hbar}}H\left(\frac{E}{\hbar
\omega}-\frac{1}{2},\sqrt{\frac{m \omega}{\hbar}}x \right),
\end{equation}
where $H(\nu,x)$ is the Hermite polynomial for integer $\nu$. It
follows then that
\begin{equation}
\psi_s(0)=2^{\frac{E}{\hbar
\omega}-\frac{1}{2}}\frac{\sqrt{\pi}}{\Gamma(\frac{3}{4}-\frac{E}{\hbar
\omega})}
\end{equation}
and
\begin{equation}
\psi_a'(0)=-2^{\frac{E}{\hbar \omega}}\sqrt{\frac{2\pi\omega m
}{\hbar}}\frac{1}{\Gamma(\frac{1}{4}-\frac{E}{\hbar \omega})}.
\end{equation}
Moreover, in the thermodynamic limit $L\rightarrow\infty$,
\begin{equation}
 G(0,0,E)=\frac{1}{2\hbar}\sqrt{\frac{m }{\omega
\hbar}}\frac{\Gamma(\frac{1}{4}-\frac{E}{\hbar
\omega})}{\Gamma(\frac{3}{4}-\frac{E}{\hbar
\omega})}.\label{HarmFSol}
\end{equation}
The Eq.\ (\ref{HarmFSol}) approaches asymptotically the expression
for free fermions, as $\omega \rightarrow 0$ for $E<0$,
\begin{equation}
 G(0,0,E)\rightarrow \frac{1}{\hbar}\sqrt{\frac{-m }{2
E}}.\label{FreeFSol}
\end{equation}

\end{document}